\documentclass[pra,aps,groupedaddress,floatfix,twocolumn,superscriptaddress,showpacs]{revtex4}
\usepackage{amsmath,amssymb,epsfig,bm,pifont}
\usepackage[linkcolor=black,citecolor=black,urlcolor=blue,colorlinks=true,
linktocpage=true]{hyperref}
\usepackage{graphicx}
\usepackage{epstopdf}
\usepackage{epsfig}
\usepackage{bm}
\usepackage{tabularx}

\newcommand{\beq}{\begin{equation}}
\newcommand{\eeq}{\end{equation}}
\newcommand{\bea}{\begin{eqnarray}}
\newcommand{\eea}{\end{eqnarray}}
\newcommand{\tr}{\mathrm{tr}}

\begin{document}

%%%%%%%%%%%%%%%%%%%%%%%%%%%%%%%%%%%%%%%%%%%%%
\title{A hybrid Monte Carlo approach to the entanglement entropy of interacting fermions}

\author{Joaqu\'{\i}n E. Drut}
\email{drut@email.unc.edu}
\affiliation{Department of Physics and Astronomy, University of North Carolina,
Chapel Hill, North Carolina, 27599-3255, USA}

\author{William J. Porter}
\email{wjporter@live.unc.edu}
\affiliation{Department of Physics and Astronomy, University of North Carolina,
Chapel Hill, North Carolina, 27599-3255, USA}

\begin{abstract}
The Monte Carlo calculation of R\'enyi entanglement entropies $S^{}_n$ of interacting fermions suffers from a well-known 
signal-to-noise problem, even for a large number of situations in which the infamous sign problem is absent. A few methods 
have been proposed to overcome this issue, such as ensemble switching and the use of auxiliary partition-function ratios. 
Here, we present an approach that builds on the recently proposed free-fermion decomposition method; it incorporates 
entanglement in the probability measure in a natural way; it takes advantage of the hybrid Monte Carlo algorithm 
(an essential tool in lattice quantum chromodynamics and other gauge theories with dynamical fermions);
and it does not suffer from noise problems. This method displays no sign problem for the same cases as other 
approaches and is therefore useful for a wide variety of systems. As a proof of principle, we calculate
$S_2^{}$ for the one-dimensional, half-filled Hubbard model and compare with results from exact diagonalization
and the free-fermion decomposition method.
\end{abstract}

\date{\today}
\maketitle

%%%%%%%%%%%%%%%%%%%%%%%%%%%%%%%%%%%%%%%%%%%%%
\section{Introduction} 

Topological and quantum-information aspects of condensed matter physics, broadly defined to include
all few- and many-body quantum systems, continue to gain increasing attention from a variety of angles, with 
the quantum-mechanical notion of entanglement playing a central role~\cite{RevModPhys}. Topological quantum phase transitions, 
in particular, have been shown to bear a direct quantitative connection to the so-called entanglement entropy in both its 
R\'enyi and von Neumann forms, the entanglement spectrum, and other information-related quantities~\cite{Kitaev}.
Thus, the computation of R\'enyi entanglement entropies $S^{}_n$ is currently of vital importance to many fields 
(see e.g.~\cite{Srednicki,CalabreseCardy,HEP}), 
and the challenge of doing so in interacting systems, particularly in strongly coupled regimes, must be met.

To this end, a variety of Monte Carlo (MC) methods have recently been put forward to calculate $S^{}_n$ efficiently 
(see e.g.~\cite{Melko, Humeniuk, McMinis, Luitz, Grover, Broecker, WangTroyer, Assaad}).
As we explain below, one of the crucial steps common to many of the underlying formalisms is the so-called 
replica trick~\cite{CalabreseCardy}, which results in 
an expression for $S^{}_n$ that consists of a ratio of two partition functions. Generally speaking, partition functions themselves 
are challenging objects to compute from the numerical standpoint, as they typically involve terms that vary 
on vastly different numerical scales. 
In the context of stochastic calculations of $S^{}_n$, it is now well understood that this complication manifests 
itself as a signal-to-noise problem: the direct estimation of the partition functions, followed by the calculation of their ratio, 
leads to statistical uncertainties that grow exponentially with the volume of the (sub-)system considered 
(see e.g.~\cite{Grover,Broecker,WangTroyer} for an explanation).

In this work, we present an alternative lattice MC approach for the calculation of $S^{}_n$. We use a 
specific case of one-dimensional spin-1/2 fermions governed by the Hubbard Hamiltonian as an example, 
which allows us to compare our results with the exact solution as well as with other MC methods, but the technique can be 
generalized to arbitrary systems, including those with gauge fields and Fermi-Bose mixtures (as long as the 
so-called sign problem is absent, as in any other MC calculation; see e.g.~\cite{MCReviews}). To highlight the 
generality of our technique, we carry out our calculations using the hybrid Monte Carlo algorithm (HMC)~\cite{HMC} 
(see~\cite{BasicHMC} for basic introductions to HMC), which 
is the workhorse of lattice QCD, it is essential in non-perturbative studies of gauge theories 
with dynamical fermions and, more recently, has been used in a variety of graphene 
studies~\cite{grapheneDL, grapheneHS, grapheneBP, grapheneBRS}. 

%%%%%%%%%%%%%%%%%%%%%%%%%%%%%%%%%%%%%%%%%%%%%%
\section{Formalism}

For the following discussion, we put the system on a $d$-dimensional spatial 
lattice of side $N_x^{}$. Because we are considering such a finite lattice, 
the single-particle space is of finite size $N_x^d$. 
We then follow closely the formalism of Ref.~\cite{Grover}.

The $n$-th R\'enyi entanglement entropy $S^{}_{n}$ of a sub-system $A$ of a
given quantum system is defined by
\beq
\label{Eq:SnDef}
S^{}_{n}= \frac{1}{1-n} \ln \tr (\hat \rho^n_A),
\eeq
where $\hat  \rho^{}_A$ is the reduced density matrix of sub-system $A$ 
(i.e. the degrees of freedom of the rest of the system are traced over).  More concretely, for a system with density matrix $\hat \rho$, the reduced density matrix is defined via a partial trace over the Hilbert space corresponding to the complement $\bar{A}$ of our sub-system as
\beq
\label{Eq:RedDM}
\hat{\rho}^{}_{A} = \tr_{\bar{A}}\hat{\rho}.
\eeq
In Ref.~\cite{Grover}, Grover derived an auxiliary-field path-integral form for
$\hat \rho^{}_A$, from which he showed that $S^{}_{n}$ can be computed using MC methods 
for a wide variety of systems. We summarize those derivations next.
In auxiliary-field Monte Carlo methods one introduces a Hubbard-Stratonovich field $\sigma$ that
decouples the fermion species, such that the usual density matrix $\hat \rho$ can be written as a path integral:
\beq
\hat \rho = \frac{e^{-\beta \hat H}}{\mathcal Z} = \int \mathcal D \sigma^{} P[\sigma]\, \hat \rho[\sigma],
\eeq
for some normalized probability measure $P[\sigma]$ determined by the details of the underlying Hamiltonian (for more detail, see~\cite{MCReviews}). 
Here, $\mathcal Z = \tr[e^{-\beta \hat H}]$ is the partition function, and $\hat \rho[\sigma]$ is an auxiliary density matrix corresponding to 
noninteracting particles in the external field $\sigma$.
In analogy with this, Grover proved that one may decompose the reduced density matrix as
\beq
\label{Eq:rhoA}
\hat{\rho}^{}_{A} = \int \mathcal D\sigma^{}P[\sigma]\,\hat{\rho}^{}_{A}[\sigma],
\eeq
where
\beq
\hat{\rho}^{}_{A}[\sigma] = C_{A}[\sigma]\; \exp\left(-\sum_{i,j} \hat c^{\dagger}_i [\ln(G^{-1}_{A}[\sigma^{}]-\openone)]_{ij}^{} \hat c^{}_j\right),
\eeq
and
\beq
C_{A}[\sigma] = \det(\openone - G^{}_{A}[\sigma^{}]).
\eeq

Here, we have used the {\it restricted Green's function} $\left [G^{}_{A}[\sigma^{}]\right ]_{ij}$, which corresponds to a noninteracting 
single-particle Green's function $G(i,j)$ in the background field $\sigma^{}$
but such that the arguments $i,j$ only take values in the region $A$ 
(see Ref.~\cite{Grover} and also Ref.~\cite{Peschel}, where expressions for the reduced density matrix
of noninteracting systems, based on reduced Green's functions, were first derived).

Using the above judicious choice of $\hat{\rho}^{}_{A}[\sigma]$, Grover verified that the expectation value of $c_j c^{\dagger}_i$ 
in the auxiliary noninteracting system reproduces the restricted single-particle Green's functions, as required.
Therefore expectation values of observables supported in the region $A$ are reproduced as well. This validates the
expression on the right-hand side of Eq.~(\ref{Eq:rhoA}). 

Using that expression, taking powers of $\hat{\rho}^{}_{A}$ results in the appearance of multiple auxiliary fields, which we will
denote below collectively as $\{\sigma\}$. Seemingly an explicit manifestation of the replica trick~\cite{CalabreseCardy}, this approach 
allows the trace of the $n$-th power of $\hat{\rho}^{}_{A}$ in Eq.~(\ref{Eq:SnDef}) to be recast as a multiple field integral over a product of
fermion determinants that depend collectively on all the $\{\sigma\}$.
Indeed, for a system of $2N$-component fermions, using a Hubbard-Stratonovich transformation that decouples all of them, we obtain
\bea
\label{Eq:SnMC}
\exp\text{\big(}(1-n)S^{}_{n} \text{\big)} =
\int \mathcal D {\{\sigma\}}^{} P[\{\sigma \}]\,Q[\{\sigma \}],
\eea
where the field integration measure, given by 
\beq
\mathcal D {\{\sigma\}}^{}  = 
\prod_{k=1}^{n} \frac{\mathcal D {\sigma^{}_k}}{\mathcal Z},
\eeq
is over the $n$ ``replicas'' $\sigma^{}_k$ of the 
Hubbard-Stratonovich auxiliary field.
For convenience, we have included the normalization
\beq
\mathcal Z = 
\int \mathcal D {\sigma}^{} \prod_{m=1}^{2N} {{\det}\,U^{}_{m}[\sigma]}
\eeq
in the integration measure. The naive probability measure, given by
\beq
\label{Eq:PP}
P[\{\sigma \}] = \prod_{k=1}^{n} \prod_{m=1}^{2N} {{\det}\,U^{}_{m}[\sigma^{}_k]},
\eeq
factorizes entirely across replicas, and is therefore blind to entanglement. 
This factorization is the main reason why using $P[\{\sigma \}]$ as a MC probability 
leads to (seemingly) insurmountable signal-to-noise issues, as shown in Ref.~\cite{Grover}; it is also why we
call it naive (although that is by no means our judgement of Ref.~\cite{Grover}). 
In Eq.~(\ref{Eq:PP}), $U_m[\sigma]$ is a matrix which encodes the dynamics of the $m$-th
component in the system, namely the kinetic energy and the form of the interaction after a
Hubbard-Stratonovich transformation; it also encodes the form of the trial state $|\Psi \rangle$
in ground-state approaches (see e.g. Ref.~\cite{MCReviews}), as is the case in this work. 
We will take $|\Psi \rangle$ to be a Slater determinant. In finite-temperature approaches, $U_m[\sigma]$ 
is obtained by evolving a complete set of single-particle states in imaginary time.

The quantity that contains the essential contributions to entanglement is 
\beq
\label{Eq:QQ}
Q[\{\sigma \}] = \prod_{m=1}^{2N} {{\det}\,M^{}_{m}[\{\sigma \}]},
\eeq
where
\bea
M_{m}[\{\sigma\}] &\equiv& \prod_{k=1}^{n} \left(\openone - G^{}_{A,m}[\sigma^{}_k]\right)\times \nonumber \\ 
&&\left[\openone + \prod_{k=1}^{n}\frac{G^{}_{A,m}[\sigma^{}_k]}{\openone - G^{}_{A,m}[\sigma^{}_k]} \right].
\eea
In the above equation, we have used $G^{}_{A,m}[\sigma^{}_k]$, which is a restricted Green's function, as previously
defined, but where we now indicate the fermion component $m$ and replica field index $k$.

The product $Q[\{\sigma \}]$ was identified as playing the role of an observable in Ref.~\cite{Grover},
which is a natural interpretation in light of Eq.~(\ref{Eq:SnMC}), but which we will interpret 
differently below. Note that, for $n\!=\!2$, no matrix inversion is
required in the calculation of $Q[\{\sigma \}]$; for higher $n$, however, there is no obvious way to avoid
the inversion of $\openone - G^{}_{A,m}[\sigma^{}_k]$. In turn, this would require some kind of numerical
regularization technique (see Ref.~\cite{Assaad}) to avoid the singularities in $G^{}_{A,m}[\sigma^{}_k]$,
whose eigenvalues can be very close to 0 and 1.

In ground-state approaches, the size of $U_m[\sigma]$ is given by the number of particles of
the $m$-th species present in the system. In finite-temperature approaches, the size is
that of the whole single-particle Hilbert space (i.e. the size of the lattice).
The size of $G^{}_{A,m}[\sigma^{}_k]$, on the other hand, is always
given by the number of lattice sites enclosed by the region $A$.
Note that, separating a factor of $\mathcal Z^n$ in the denominator of
Eq.~(\ref{Eq:SnMC}), an explicit form can be identified in the numerator as the result of the so-called replica
trick~\cite{CalabreseCardy} (namely a partition function for $n$ copies of the system, ``glued'' together in the region $A$).

%%%%%%%%%%%%%%%%%%%%%%%%%%%%%%%%%%%%%%%%%%%%%%
\section{Our proposed method}

In analogy to the right-hand side of Eq.~(\ref{Eq:SnMC}), we introduce an auxiliary 
parameter $0\le\lambda \le 1$ and define a function $\Gamma(\lambda;g)$ via 
\beq
\label{Eq:GammaDef}
\Gamma(\lambda;g) \equiv \int \mathcal D {\{\sigma\}}^{} P[\{\sigma \}]\;
Q[\{\sigma \};\lambda],
\eeq
where we have replaced the dependence of $Q[\{\sigma\}]$ on the coupling $g$ by setting
\beq
g \to \lambda^2 g,
\eeq
which defines $Q[\{\sigma \};\lambda]$.
Using this definition, it follows immediately that $\Gamma(\lambda;g)$ satisfies two 
important constraints with physical significance: For $\lambda=0$, we find
\beq
\frac{1}{1-n}\ln \Gamma(0;g)=S_{n}^{(0)},
\eeq
where $S_{n}^{(0)}$ corresponds to a noninteracting system. Indeed, at $\lambda=0$ the quantity 
$Q[\{\sigma \};\lambda]$ does not depend on $\{\sigma\}$
and factors out of the integral. Thus, regardless of the value of $g$, the function $\Gamma(0;g)$ corresponds
to the R\'enyi entropy of a noninteracting system, which can be trivially computed with the present formalism.
Indeed, there is no path integral when interactions are turned off, such that the noninteracting result can be computed
with a single Monte Carlo sample using the formalism by Grover mentioned above.
It is worth noting at this point that the R\'enyi entropy of a noninteracting system has 
received substantial attention in the last few years. Much is known about this quantity
for a variety of systems, in particular in connection with area laws and their violation~\cite{FreeGasEE}.

For $\lambda = 1$, on the other hand, $\Gamma(\lambda;g)$ yields
the entanglement entropy of the fully interacting system:
\beq
\frac{1}{1-n}\ln \Gamma(1;g)=S_{n}^{}.
\eeq
Thus, both of these reference points are physically meaningful, one
of them is comparatively trivial to obtain, and obtaining the other one is our objective.

Using Eq.~(\ref{Eq:GammaDef}), 
\beq
\label{Eq:dlnGammadlambda}
\frac{\partial \ln \Gamma}{\partial \lambda}=
\int \mathcal D {\{\sigma\}}^{} \tilde P[\{\sigma \};\lambda]\; \tilde{Q}[\{\sigma \};\lambda]
\eeq
where
\beq
\label{Eq:Ptilde}
\tilde{P}[\{\sigma \};\lambda]=\frac{1}{\Gamma(\lambda;g)}P[\{\sigma \}]\;Q[\{\sigma \};\lambda],
\eeq
and
\beq
\label{Eq:Qlambda}
\tilde{Q}[\{\sigma \};\lambda]=\sum_{m=1}^{2N}\tr\left[M_{m,\lambda}^{-1}[\{\sigma \}]
\frac{\partial M_{m,\lambda}^{}[\{\sigma \}]}{\partial \lambda}\right].
\eeq
Crucially, the dependence on the parameter $\lambda$ enters only
through the matrix $M^{}_n$, and it is in this way that we propose to include the
entanglement-related correlations in the sampling procedure, which
is to be carried out using $\tilde{P}[\{\sigma \};\lambda]$ as a probability measure.
When an even number $2N$ of flavors is considered, and the interactions are attractive, 
$\det^{2N} U[\sigma]$ and $Q[\{\sigma\},\lambda]$ are real and positive semidefinite for all $\sigma$, which means 
that there is no sign problem and $\tilde P[\{\sigma \};\lambda]$ above is indeed a well-defined, normalized probability measure.

More concretely, our proposal to calculate $S^{}_n$ is to take the noninteracting $\lambda=0$ point
as a reference and compute $S_n$ using
\beq
\label{Eq:SnMCFinal}
S_{n}^{} = S_{n}^{(0)} +\frac{1}{1-n}
\int_{0}^{1}d\lambda\;\langle\tilde{Q}[\{\sigma \};\lambda]\rangle,
\eeq
where
\beq
\langle X\rangle=\int \mathcal D {\{\sigma\}}^{} \tilde{P}[\{\sigma\};\lambda]\; X[\{\sigma \}].
\eeq
In other words, we obtain an integral form of the interacting R\'enyi entropy
that can be computed using any MC method, in particular HMC~\cite{HMC}. 
The latter combines molecular dynamics (MD) of the auxiliary fields (defining a fictitious auxiliary 
conjugate momentum) with the Metropolis algorithm, and thus enables
simultaneous global updates of all the auxiliary fields $\{ \sigma \}$. As it well known, 
HMC is a highly efficient sampling strategy, particularly when gauge fields are involved (see e.g.~\cite{HMC,MCReviews}).
The integration of the MD equations of motion requires the calculation of the MD force, 
which is given by the functional derivative of $\tilde{P}[\{\sigma\};\lambda]$ with respect to $\{\sigma\}$,
which can be calculated from Eq.~(\ref{Eq:Ptilde}).

Our proposal is akin to the so-called coupling-constant integration approach of many-body physics, but it
differs in that we have strategically introduced the $\lambda$ dependence only in the
entanglement-sensitive determinant $Q[\{\sigma \};\lambda]$ of Eq.~(\ref{Eq:QQ}).

Equation~(\ref{Eq:SnMCFinal}) is our main result and defines our method.  An essential point is
that the expectation that appears above is taken with respect to the probability
measure $\tilde{P}[\{\sigma \};\lambda]\;$, which incorporates the correlations that account 
for entanglement. In stark contrast to the naive MC probability $P[\{\sigma \}]$, which 
corresponds to statistically independent copies of the Hubbard-Stratonovich field, this
distribution does not display the decoupling responsible for the signal-to-noise
problem mentioned above.

In practice, using Eq.~(\ref{Eq:SnMCFinal}) requires MC calculations to evaluate 
$\langle\tilde{Q}[\{\sigma \};\lambda]\rangle$ as a function of $\lambda$, followed by numerical integration over $\lambda$. 
We find that $\langle\tilde{Q}[\{\sigma \};\lambda]\rangle$ is a smooth function of $\lambda$ that vanishes at $\lambda=0$
and presents most of its features close to $\lambda=1$ (further details below). We therefore carry out 
the numerical integration using the Gauss-Legendre quadrature method~\cite{NR}.
It should also be pointed out that the stochastic evaluation of $\langle\tilde{Q}[\{\sigma \};\lambda]\rangle$, for
fixed subregion $A$, can be expected to feature roughly symmetric fluctuations about the mean. Therefore,
after integrating over $\lambda$, the statistical effects on the entropy are reduced (as we show empirically
in the Results section).

A few remarks are in order regarding the auxiliary parameter $\lambda$. First, we could have performed the 
replacement $g \to \lambda^2 g$ everywhere, i.e. not only in $Q$ but also in $P$ (and its normalization $\mathcal Z$). 
This would have led to three terms in the derivative of $PQ$ with respect to $\lambda$, two of which would
come from $P$ (recall $P$ is normalized) and feature different signs and a rather indirect connection to $S_n^{}$ 
(recall $P$ factorizes across replicas). Our approach avoids this extra complication by focusing on the 
entanglement part of the path integral (i.e. $Q$). Second, we could have used $\lambda^x$ instead of $\lambda^2$,
where $x$ does not have to be an integer (although it would be rather inconvenient to make it less than 1).
This is indeed a possibility and it allows for further optimization than pursued here. In the remainder of this work
we set $x=2$, as above. Finally, the required calculations for different values of $\lambda$ are completely
independent from one another and can therefore be performed in parallel with essentially perfect scaling 
(up to the final data gathering and quadrature).

%%%%%%%%%%%%%%%%%%%%%%%%%%%%%%%%%%%%%%%%%%%%%%
\section{Relation to other methods}

Our approach is very similar to the temperature-integration method of Ref.~\cite{Melko}, 
but is closer in nature to the ratio trick (and similar) of Refs.~\cite{Humeniuk, WangTroyer}.
As above, the calculation starts from the replica trick of Calabrese and Cardy~\cite{CalabreseCardy}, i.e.
\beq
\exp((1-n) S_n^{}) = \frac{\mathcal Z^{}_{A,n}}{\mathcal Z^n},
\eeq
where $\mathcal Z^{}_{A,n}$ is the partition function of $n$ copies of the
system ``glued'' together in the region $A$. Typically, $\mathcal Z^{}_{A,n}$
and $\mathcal Z^{n}_{}$ can be very different from each other in magnitude, particularly if $S_n^{}$ is large
(as is typically the case for large sub-system sizes).
Therefore, computing the above partition functions separately (and stochastically) and then attempting to evaluate the
ratio is likely to yield a large statistical uncertainty. 
A way around this problem is to use the ratio (or increment) trick, whereby one introduces auxiliary ratios of the 
partition function corresponding to systems whose configuration spaces are only marginally 
dissimilar. In other words, one writes 
\beq
\label{Eq:RatioTrick}
\exp((1\!-\!n) S_n^{}) = 
\frac{\mathcal Z^{}_{A,n}}{\mathcal Z^n} = 
\frac{\mathcal Z^{}_{A,n}}{\mathcal Z^{}_{A-1,n}}
\frac{\mathcal Z^{}_{A-1,n}}{\mathcal Z^{}_{A-2,n}}
\cdots
\frac{\mathcal Z^{}_{2,n}}{\mathcal Z^{}_{1,n}}
\frac{\mathcal Z^{}_{1,n}}{\mathcal Z^n},
\eeq
where the auxiliary ratios ${\mathcal Z^{}_{A-i,n}}/{\mathcal Z^{}_{A-i-1,n}}$ are chosen to 
correspond to subsystems of similar size and shape (e.g. such that their linear
dimension differs by one lattice point). In this way, each of the auxiliary ratios can be expected to
not differ significantly from unity. With enough intermediate ratios, calculations
can be carried out in a stable fashion at the price of calculating a potentially large number of
ratios.

In the method we propose here, the parameter $\lambda$ plays the role of the varying region size $A$
of the ratio trick. Indeed, using Eq.~(\ref{Eq:SnMCFinal}), we may schematically write 
\beq
\exp\left((1\!-\!n) (S_n^{} - S_n^{(0)})\right) = 
\prod_{\lambda=0}^1 \exp \left( \Delta \lambda\;\langle\tilde{Q}[\{\sigma \};\lambda]\rangle \right),
\eeq
where any discretization $\Delta \lambda$ of the exponent inside the product yields a telescopic sequence of ratios as 
in Eq.~(\ref{Eq:RatioTrick}). As long as $\langle\tilde{Q}[\{\sigma \};\lambda]\rangle$ is regular in $\lambda$, 
which we find to be the case, our auxiliary factors can be made to be arbitrarily close to unity at the cost of (at most) linear
scaling in computation time.

%%%%%%%%%%%%%%%%%%%%%%%%%%%%%%%%%%%%%%%%%%%%%%
\section{Results}

We test our algorithm by computing the second R\'enyi entropy $S_2^{}$ for
one-dimensional, ten-site, half-filled Hubbard models with periodic boundary
conditions. The Hamiltonian we used is
\beq
\hat H  = - t \sum_{s,\langle ij\rangle}
{\left(\hat c^{\dagger}_{i,s}\hat c^{}_{j,s}+\hat c^{\dagger}_{j,s}\hat c^{}_{i,s}\right)}+
U\sum_{i}{\hat n^{}_{i\uparrow}\hat n^{}_{i\downarrow}},
\eeq
where the first sum includes two fermion flavors $s = \uparrow,\downarrow$ and nearest-neighbor pairs.
To carry out our tests, we implemented a symmetric Trotter-Suzuki decomposition of the Boltzmann
weight, with an imaginary-time discretization parameter $\tau=0.05$ (in lattice units). The full length of the
imaginary-time direction was at most $\beta = 5$ (i.e. we used 100 imaginary-time lattice points).
The interaction factor in the Trotter-Suzuki decomposition was addressed, as anticipated in a previous section, 
by introducing a replica auxiliary field $\sigma$ for each power of the reduced density matrix. This 
insertion was accomplished via a Hubbard-Stratonovich transformation, which we chose to be of a continuous but compact
form (see Ref.~\cite{MCReviews}).

\begin{figure}[h]
\includegraphics[width=1.1\columnwidth]{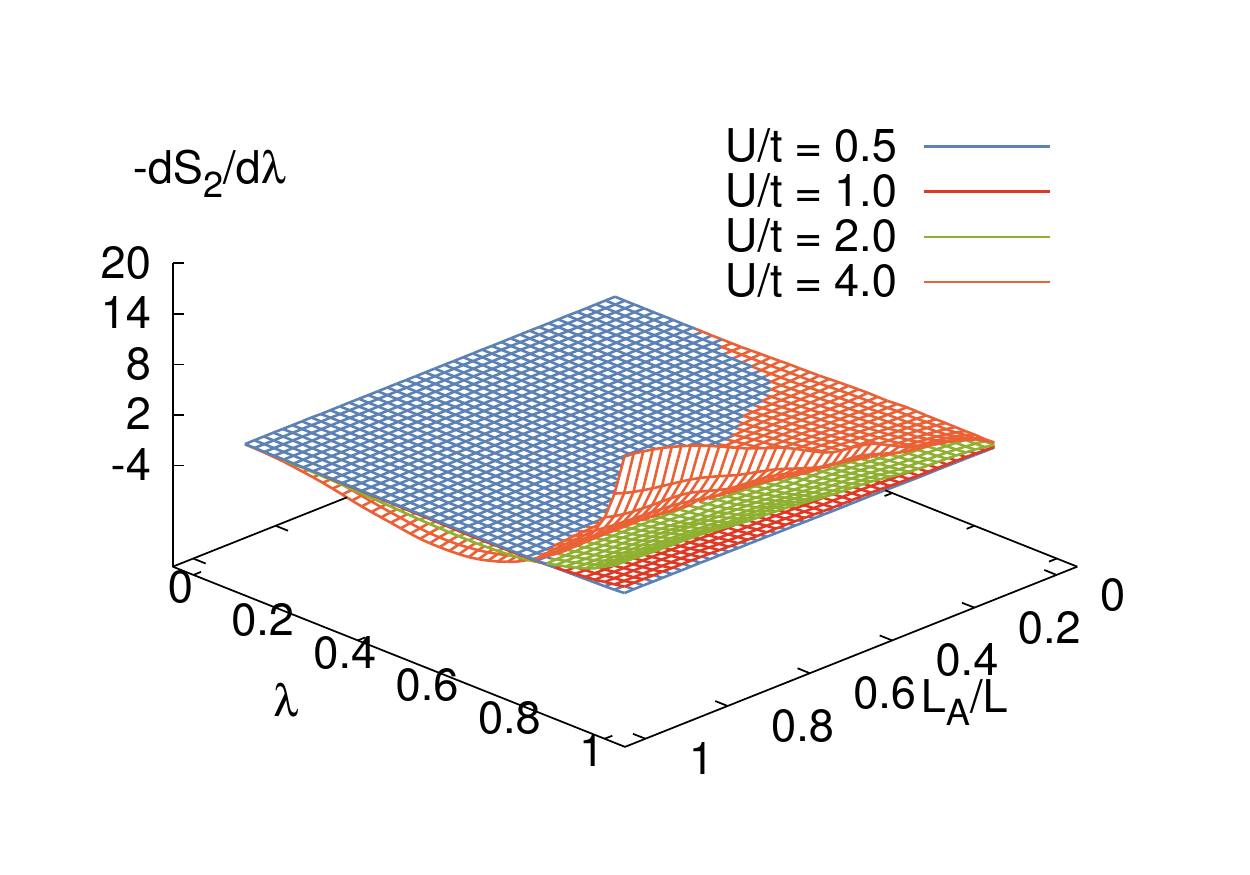}
\caption{\label{Fig:threecouplingsdsdlambda10}(color online) 
Hybrid Monte Carlo results for the source with $U/t =0.5, 1.0, 2.0$ and $4.0$ 
(from bottom to top at $\lambda=1$)
as functions of the parameter $\lambda$ and the region size $L_A$, for $N_x^{}=10$ sites.}
\end{figure}

Figure~\ref{Fig:threecouplingsdsdlambda10} plots $\langle\tilde{Q}[\{\sigma \};\lambda]\rangle$ 
as a function of both $\lambda$ and the subregion size $L_A$ for four values 
of the coupling.  We note that surfaces corresponding to weak couplings show much less fluctuation in 
both parameters than their strong-coupling counterparts.  This uniformity implies that for weakly 
coupled systems, even at large $L_A$, a coarse $\lambda$ discretization may yield good estimates 
of $S^{}_n$. Conversely, strongly coupled systems are, not unexpectedly, more computationally demanding.

For small subsystems, in addition to relatively small variation, the majority of the deviation from the 
noninteracting entropy is accumulated at large $\lambda$ and appears with mostly uniform signature. 
Much to the contrary, for larger subsystems, intermediate values of $\lambda$ correspond to a region 
of parameter space that contributes opposite-sign corrections to the entropy, which yields increasing 
uncertainty as a function of the subsystem size. This effect is most clearly seen for the curve corresponding 
to the case where the subregion constitutes the entire system. In that case, $S^{}_2$ is 
zero regardless of the coupling, which means that the integral over $\lambda$ must vanish. This happens 
by a precise cancelation that must be captured by the numerical integration procedure. Given that 
the features of $\langle\tilde{Q}[\{\sigma \};\lambda]\rangle$ are roughly concentrated in $0.5<\lambda<1$, we chose 
the Gauss-Legendre quadrature method to carry out the integral in a precise fashion. Using $N_\lambda^{}=20$ points
in the interval $[0,1]$ (i.e. 40 points in the defining interval $[-1,1]$ using an even extension of the integrand), we 
find that, for the parameter values studied here, the systematic effects associated with $\lambda$ are
smaller than the statistical uncertainty. 

Our experience, as detailed above, indicates that the features of $\langle\tilde{Q}[\{\sigma \};\lambda]\rangle$
are generic: they vary in amplitude with the coupling but are largely insensitive to the overall system size,
and generally behave in a benign way as a function of $\lambda$ and the sub-system size.
Therefore the $\lambda$ integration does not contribute to the scaling of the 
computation time vs. system size beyond a prefactor. Next, we present our results upon integrating over 
$\lambda$ as detailed above.

%%%%%%%%%%%%%%%%%%%%%%%%
\subsection{Comparison with exact diagonalization results and a first look at statistical effects}

\begin{figure}[h]
\includegraphics[width=1.0\columnwidth]{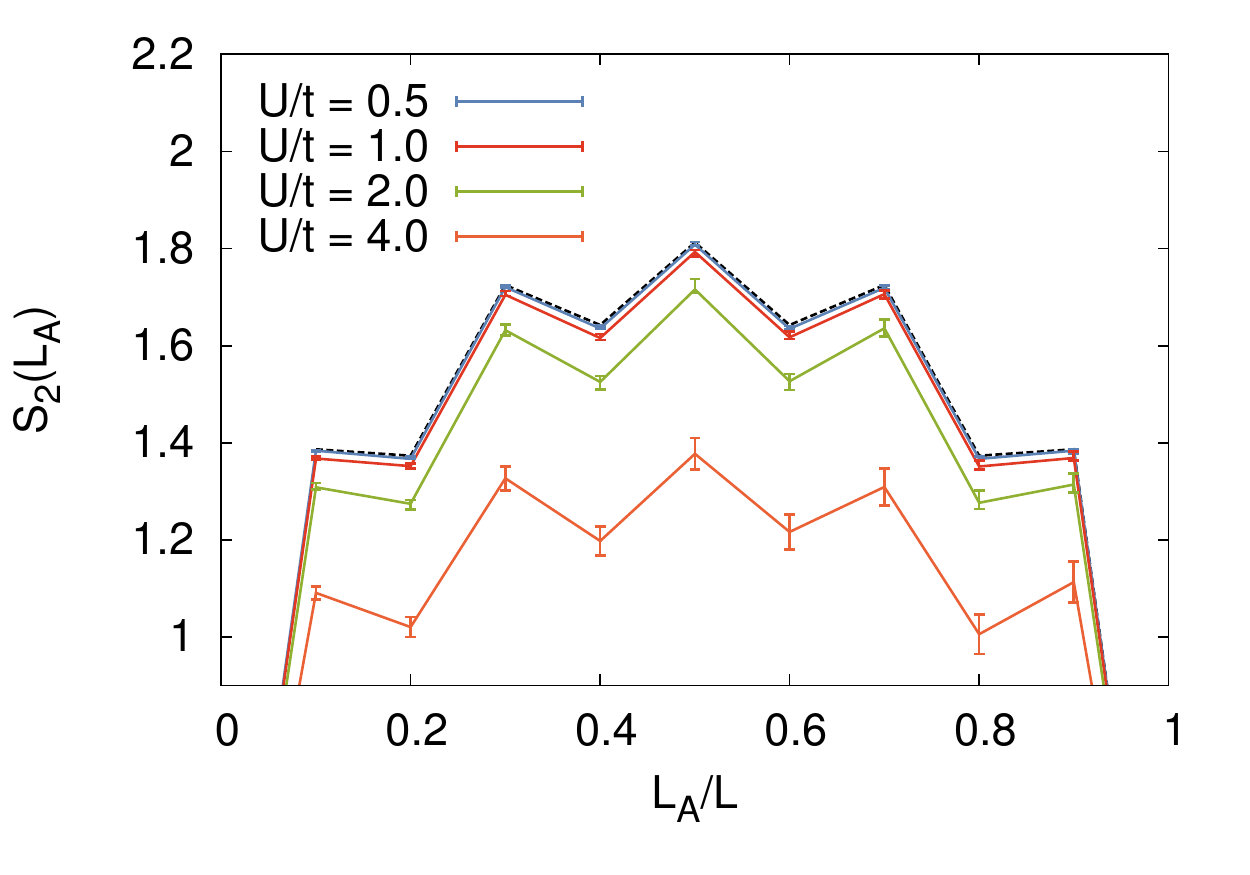}
\caption{\label{Fig:hubbard10zoom}(color online) 
Results for the ten-site Hubbard model with $U/t =0.5$, $1.0$, $2.0$, and $4.0$
(solid lines from top to bottom). The results for the noninteracting case are shown with a dashed line.
Hybrid Monte Carlo answers with numerical uncertainties for 7,500 decorrelated samples, shown with error bars. 
Exact-diagonalization results of Ref.~\cite{Grover} shown with lines, except for the $U/t =4.0$ case,
where the lines join the central values of our results and are provided to guide the eye.
}
\end{figure}

In Fig.~\ref{Fig:hubbard10zoom}, we show results for a system of size $L=N_x^{}\ell$, where $N^{}_x = 10$ sites and
$\ell=1$ is the lattice spacing (as in conventional Hubbard-model studies).
Our results cover couplings $U/t$ = 0.5, 1, 2, and 4, and subsystems of size $L^{}_A=1-10$, which we compare to 
the results of Ref.~\cite{Grover}. The agreement is quite 
satisfactory. 

\begin{figure}[t]
\includegraphics[width = 0.943\columnwidth]{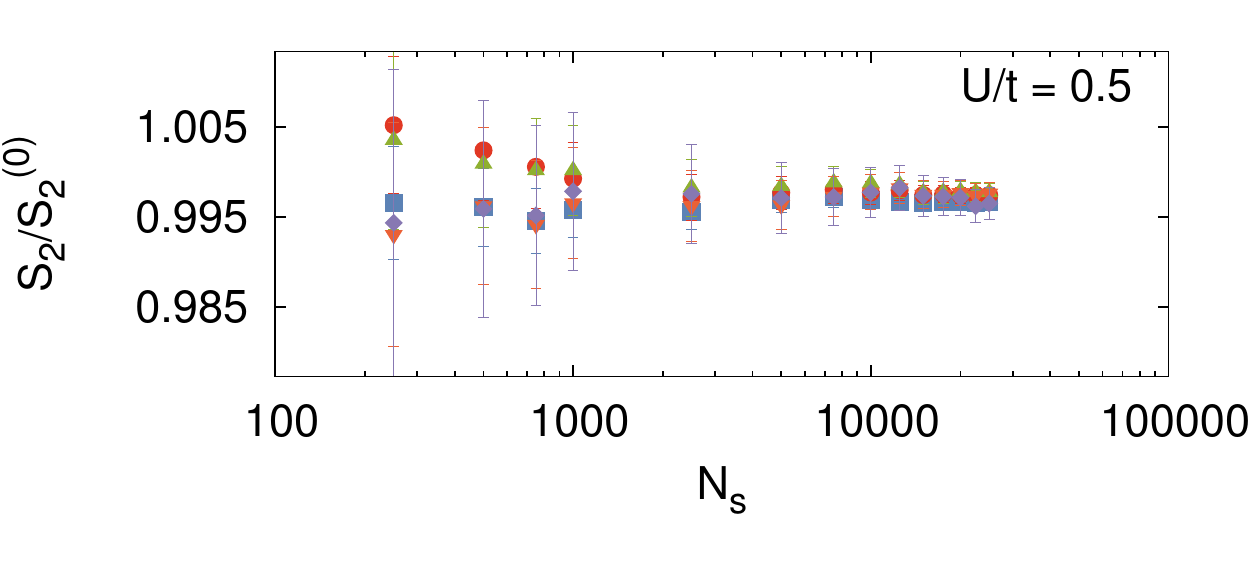}
\includegraphics[width = 0.943\columnwidth]{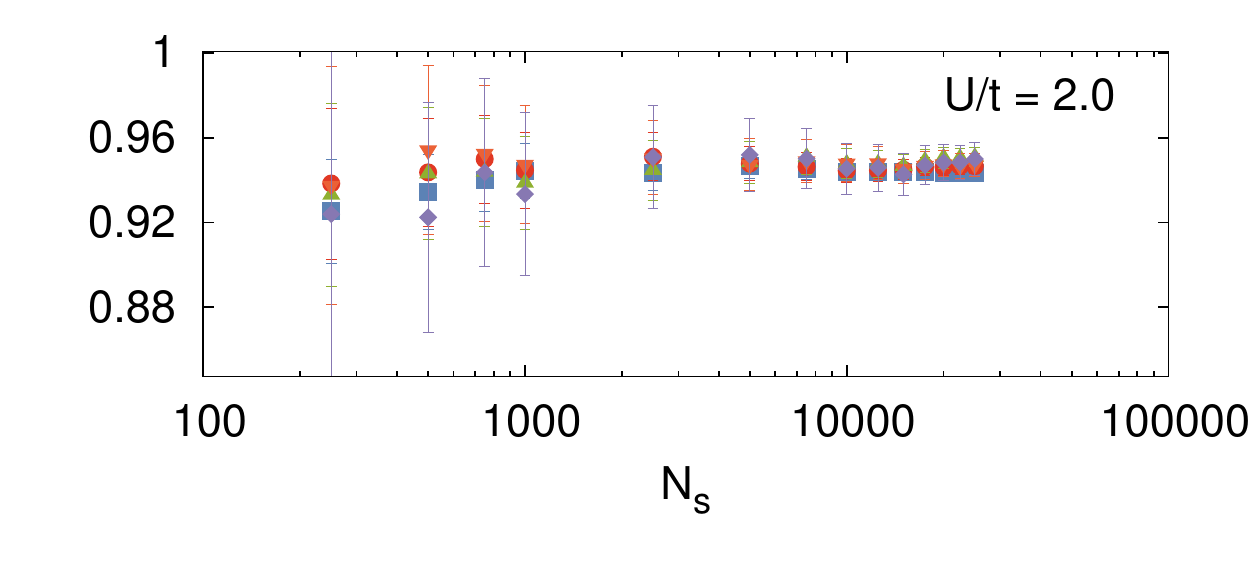}
\includegraphics[width = 0.943\columnwidth]{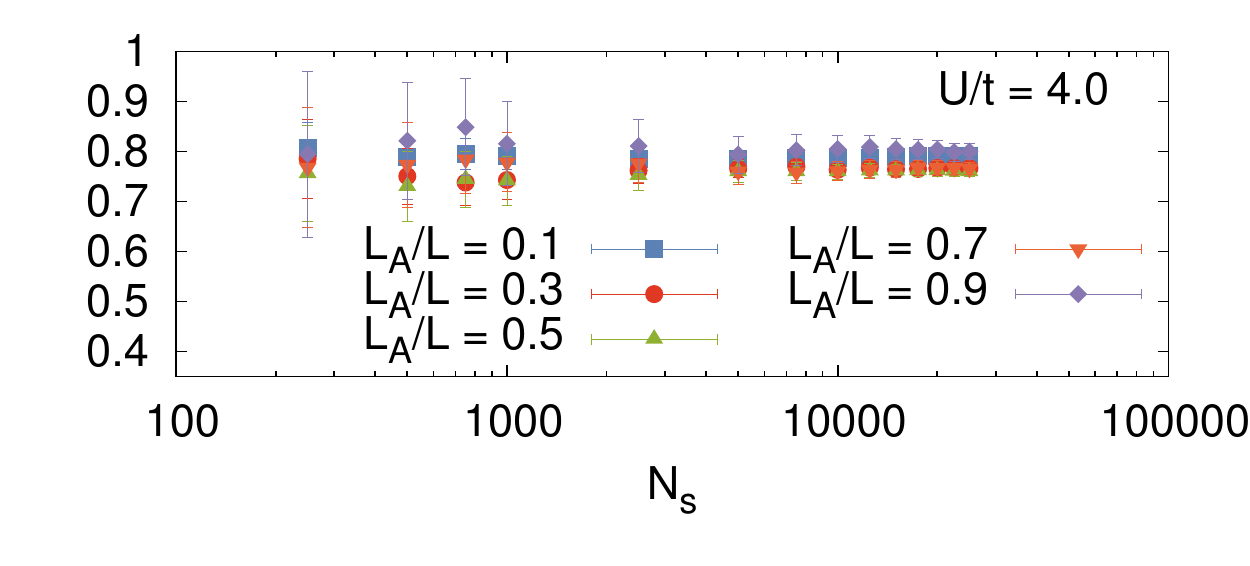}
\caption{\label{Fig:ConvergencePlotNx10}(color online) 
Second R\'enyi entropy (scaled to the noninteracting result) as a function of the number of samples 
$N^{}_{s}$ for coupling $U/t = 0.5$, $2.0$, and $4.0$ shown top to bottom. Within a few thousand samples, 
we observe that the results have stabilized to within 1-2\%.}
\end{figure}

To better understand the size of the statistical effects, we also show how our results vary 
with the number of MC samples in Fig.~\ref{Fig:ConvergencePlotNx10}. In Fig.~\ref{Fig:ConvergencePlotNx10} 
we see that the 25,000 samples collected were well beyond what was needed: half as many
would have already given excellent results. These results show that, by including entanglement-sensitive
contributions into the probability measure, our approach circumvents the signal-to-noise problem mentioned in
the introduction. Below we elaborate more explicitly on statistical effects and said problem, and show concrete 
numerical examples of how it arises in practice.
\begin{figure}[h]
\includegraphics[width = 1\columnwidth]{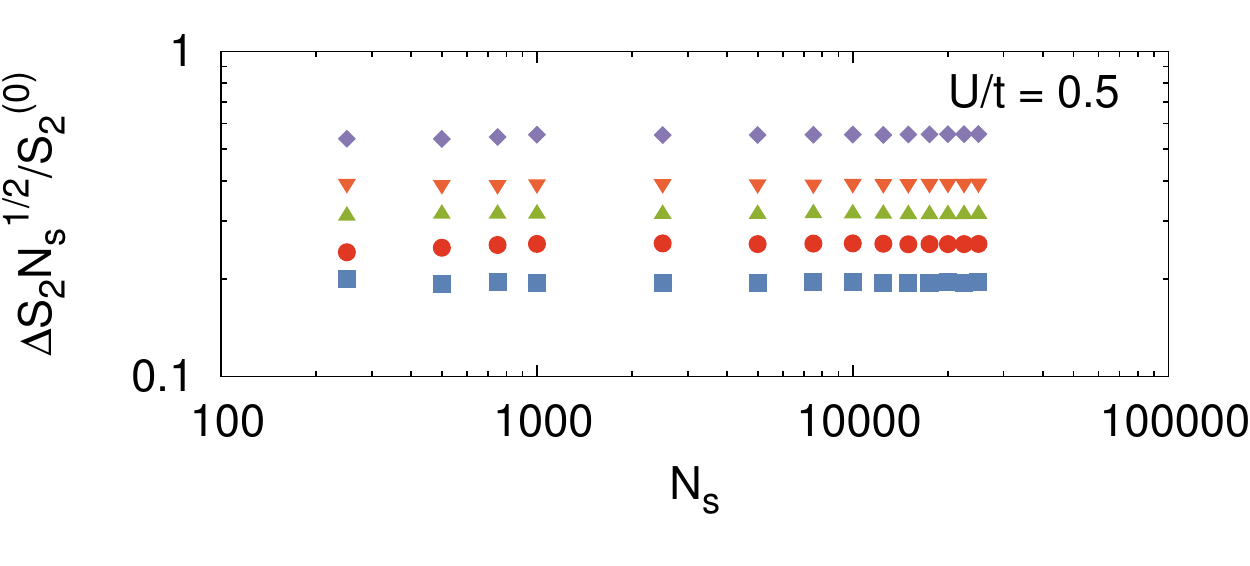}
\includegraphics[width = 1\columnwidth]{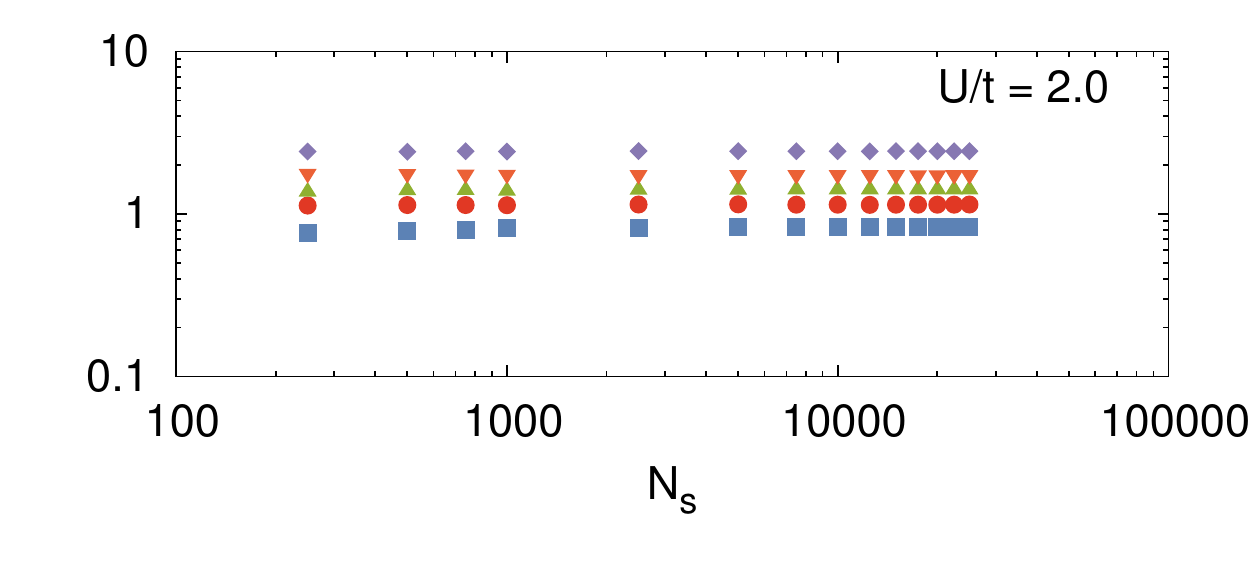}
\includegraphics[width = 1\columnwidth]{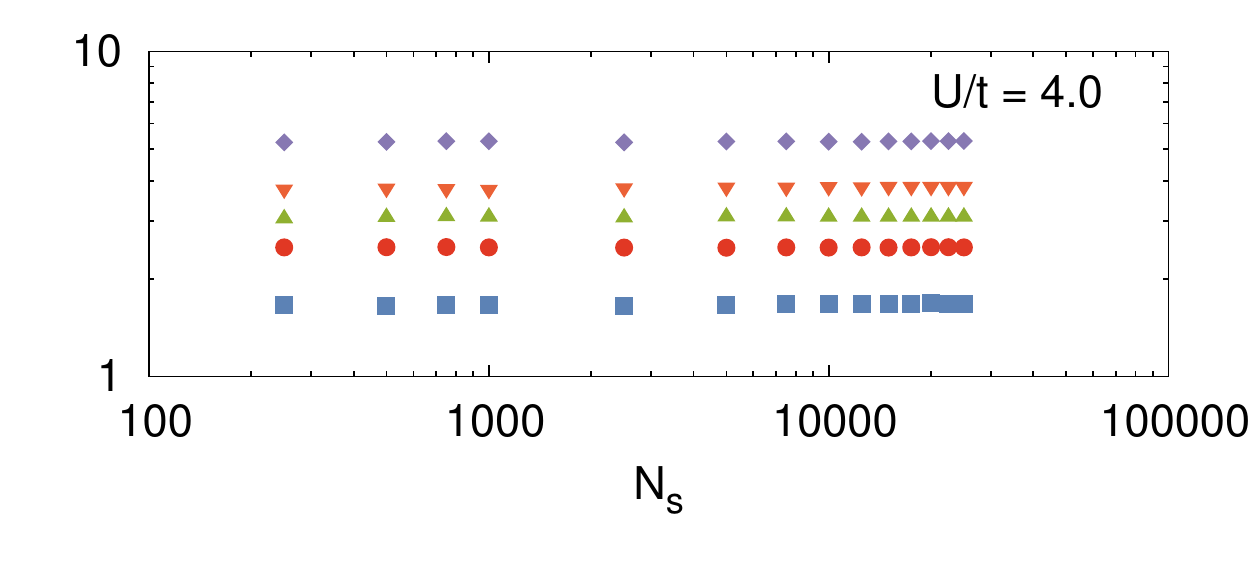}
\caption{\label{Fig:ErrorRootNsPlot}(color online) Relative statistical uncertainty of the second R\'enyi entropy 
(propagated from the standard deviation in the uncertainties on its $\lambda$ derivative) 
as a function of the number of samples $N^{}_{s}$ for couplings $U/t = 0.5$, $2.0$, and $4.0$ shown top to bottom.
The symbols and colors correspond to those utilized in Fig.~\ref{Fig:ConvergencePlotNx10}.
}
\end{figure}

In Fig.~\ref{Fig:ErrorRootNsPlot}, we show the overall statistical uncertainty $\Delta S_2^{}$ in our estimates of $S_2^{}$ as a 
function of the number of samples $N_s^{}$, for the coupling strengths and subsystem sizes studied above. 
$\Delta S_2^{}$ was computed by using the envelope determined by the MC statistical uncertainties in 
$\langle\tilde{Q}[\{\sigma \};\lambda]\rangle$ as a function of $\lambda$.
While $\Delta S_2^{}$ grows with the sub-system size, its $N_s^{1/2}$ scaling remains constant as the
number of samples is increased.

%%%%%%%%%%%%%%%%%%%%%%%%%%%%%%%%%%%%%%%%%%%%%%%%
\subsection{Comparison with naive free-fermion decomposition method}

\begin{figure}[h]
\includegraphics[width = 1\columnwidth]{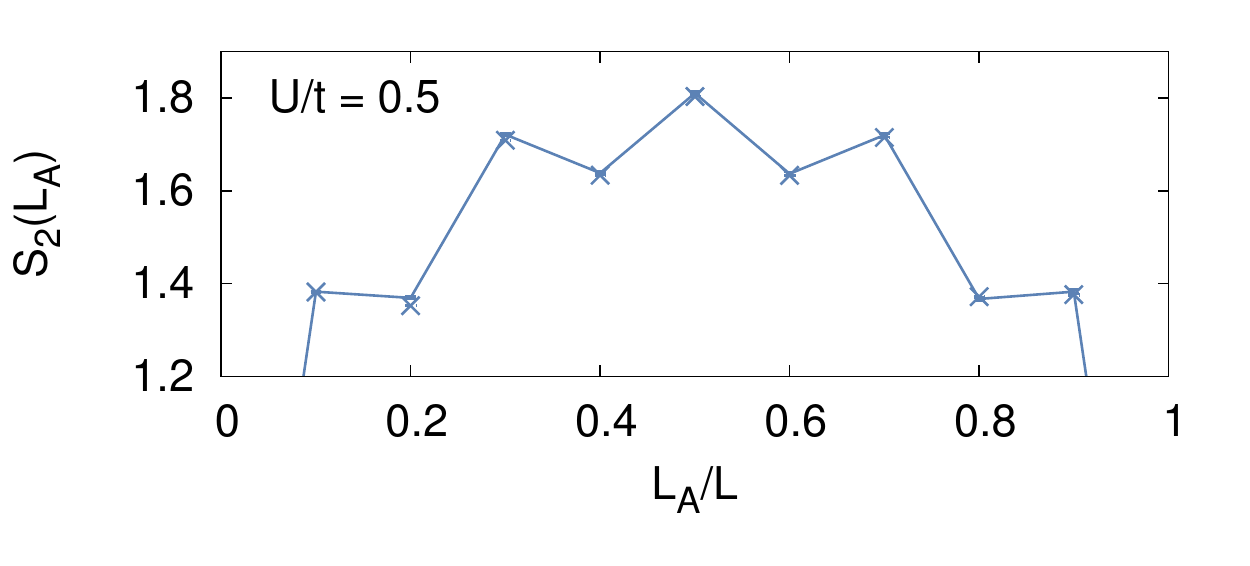}
\includegraphics[width = 1\columnwidth]{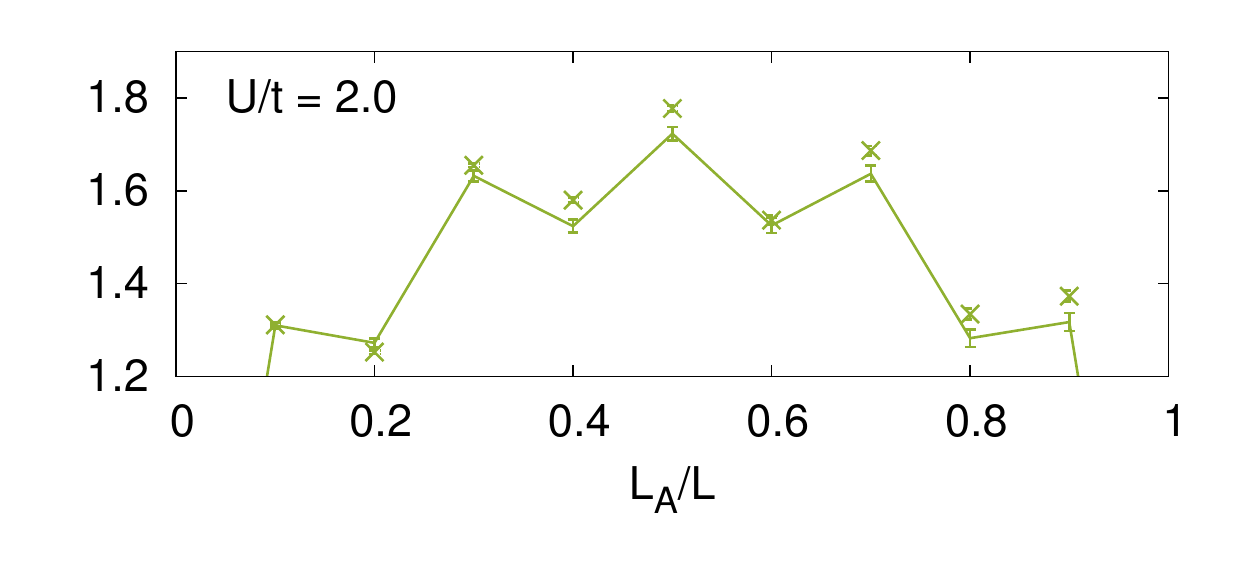}
\includegraphics[width = 1\columnwidth]{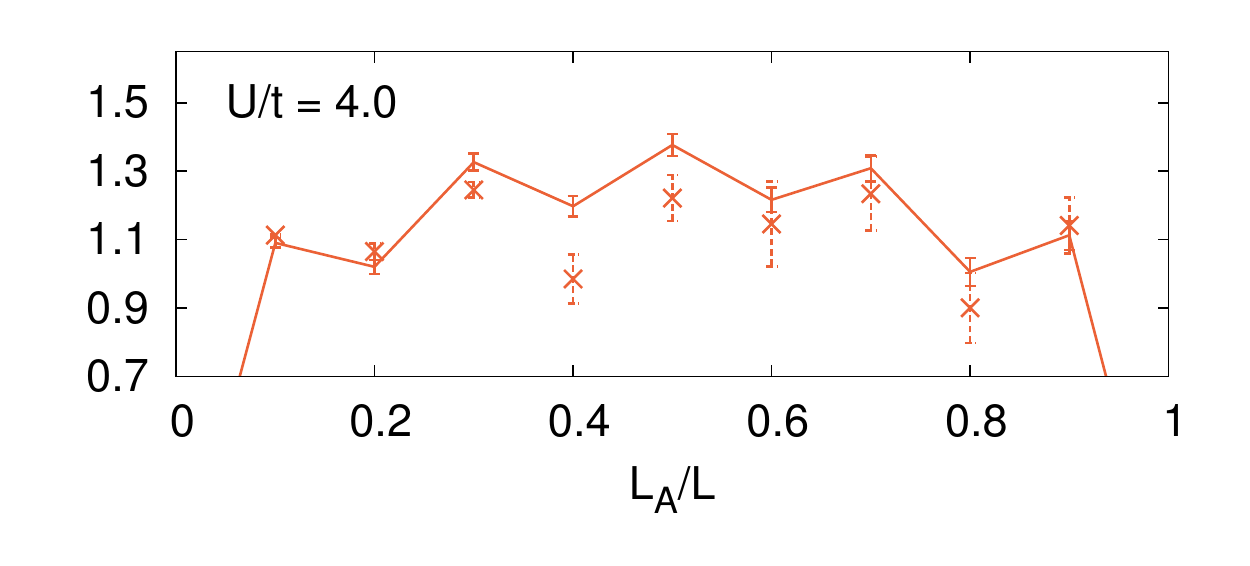}
\caption{\label{Fig:FreeFermionCompare}(color online) 
Hybrid Monte Carlo results (solid) for the ten-site Hubbard model with $U/t =0.5$, $2.0$, and $4.0$
with numerical uncertainties for 7,500 decorrelated samples (for each value of $\lambda$) compared with 
results from the naive free-fermion decomposition method (crosses with dashed error bars) with 75,000 decorrelated samples.}
\end{figure}
\begin{figure}[t]
\includegraphics[width=1.0\columnwidth]{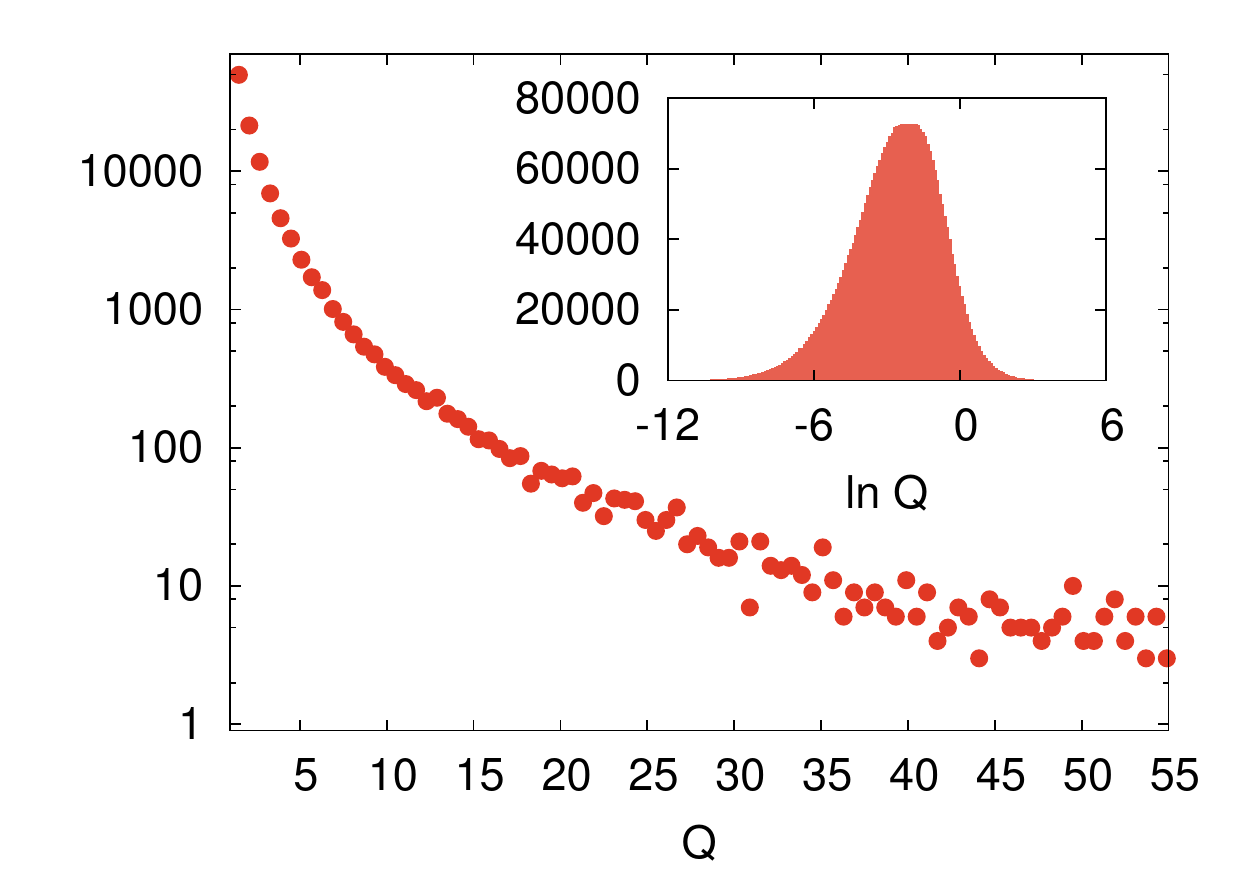}
\caption{\label{Fig:NaiveHistogram}(color online) 
Distribution of the observable $Q[\{\sigma\}]$ of the naive free-fermion decomposition method [implemented 
via Eq.~(\ref{Eq:SnMC})] for
$U/t=2.0$ and $L^{}_A/L = 0.8$. Note that $Q[\{\sigma\}]$ is a non-negative quantity.
The long tail (main plot; note logarithmic scale in vertical axis) extends beyond the range shown and 
is approximately a log-normal distribution, i.e. $\ln Q[\{\sigma\}]$ is roughly distributed as a gaussian (inset).
}
\end{figure}

Figure~\ref{Fig:FreeFermionCompare} again shows our results for the ten-site Hubbard model using 7,500 decorrelated samples
for each value of $\lambda$, this time compared with the naive free-fermion decomposition method using 75,000 samples. 
Such an increased number of samples for the naive method was chosen to provide a more fair comparison with our method,
considering that the latter requires a MC calculation for each value of $\lambda$. We used 20 $\lambda$ points but, as explained 
in more detail below, roughly half of the $\lambda$ points require only a small number of samples.

The statistical uncertainties for the naive method do not encompass the expected answers for many of the points, which is a symptom 
of an ``overlap'' problem, i.e. the probability measure employed bears little correlation with the observable, as mentioned above 
(see also Ref.~\cite{NoiseSignProblemStatistics}). This is the same issue as the signal-to-noise problem referred to above.

To illustrate this situation more precisely, we show in Fig.~\ref{Fig:NaiveHistogram} a histogram of $Q[\{\sigma \}]$ [see Eq.~(\ref{Eq:SnMC})]
for $U/t=2.0$ and $L_{A}/L=0.8$. 
Even using a logarithmic vertical scale, the distribution displays a long tail that extends across multiple orders of 
magnitude. We find that the distribution is approximately of the log-normal type 
(i.e. its logarithm is approximately distributed as a gaussian, as shown in the inset of Fig.~\ref{Fig:NaiveHistogram});
this is the challenge faced when attempting to determine the average of $Q[\{\sigma \}]$
with good precision implementing the free-fermion decomposition of Ref.~\cite{Grover} at face value. 
Moreover, we expect these features to worsen in larger systems, higher dimensions, and stronger couplings, as the matrices 
involved become more ill-conditioned.

Notably, it is the logarithm of the expectation value of $Q[\{\sigma \}]$ that determines the entanglement entropy, 
which could then be obtained using the cumulant expansion. 
However, it is a priori entirely unknown whether such an expansion would converge, i.e. we do not know to what extent 
this distribution deviates from gaussianity. 

The log-normality referred to above has been associated with the auxiliary field representation of the interaction. 
In such external fields $\sigma$, the orbitals of the trial wavefunction diffuse much like electrons in a disordered medium, and the stronger 
the interaction (or the lower the temperature) the heavier the tail becomes in the distribution of $Q[\{\sigma \}]$. 
This effect was noticed relatively recently in Ref.~\cite{NoiseSignProblemStatistics}, and it appears
to be quite ubiquitous. It was then shown, phenomenologically, that many signal-to-noise problems are characterized by 
the heavy tail of a lognormal distribution (see also, Ref.~\cite{deGrand}). 

%%%%%%%%%%%%%%%%%%%%%%%%%%%%%%%%%%%%%%%%%%%%%%%%
\subsection{Statistical behavior as a function of coupling, region size, and auxiliary parameter}

\begin{figure}[h]
\includegraphics[width=1.0\columnwidth]{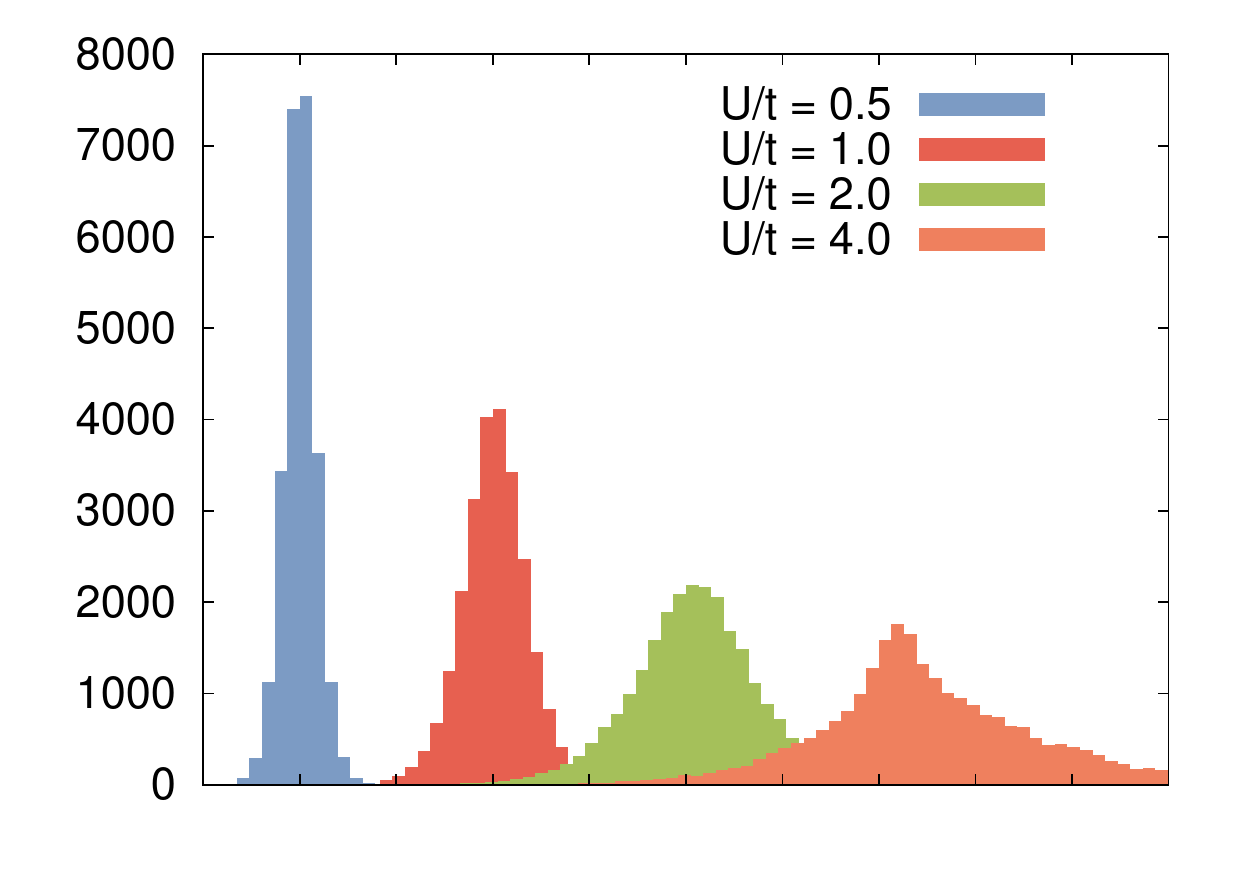}
\caption{\label{Fig:HistogramU}(color online) 
Histogram showing the statistical distribution of our results for $dS_2^{}/d\lambda$ for 
$L_A^{}/L=0.8$, $\lambda \simeq 0.83$, and $U/t =0.5, 1.0, 2.0, 4.0$. The results for 
different couplings have been shifted for display purposes, but the scale is the same
for each of them. This illustrates that, even though our method addresses the
original signal-to-noise issue, strong couplings remain more challenging
that weak couplings. 
}
\end{figure}

In Fig.~\ref{Fig:HistogramU} we show the statistical distribution of our results for the $\lambda$
derivative, for several couplings. The distributions we observe are approximately gaussian (they decay
faster than linearly in a log scale), except for the
strongest coupling we studied $U/t=4.0$ where, not unexpectedly, the distribution becomes more asymmetric
and develops heavier tails relative to its weaker-coupling counterparts. 

\begin{figure}[h]
\includegraphics[width=1.0\columnwidth]{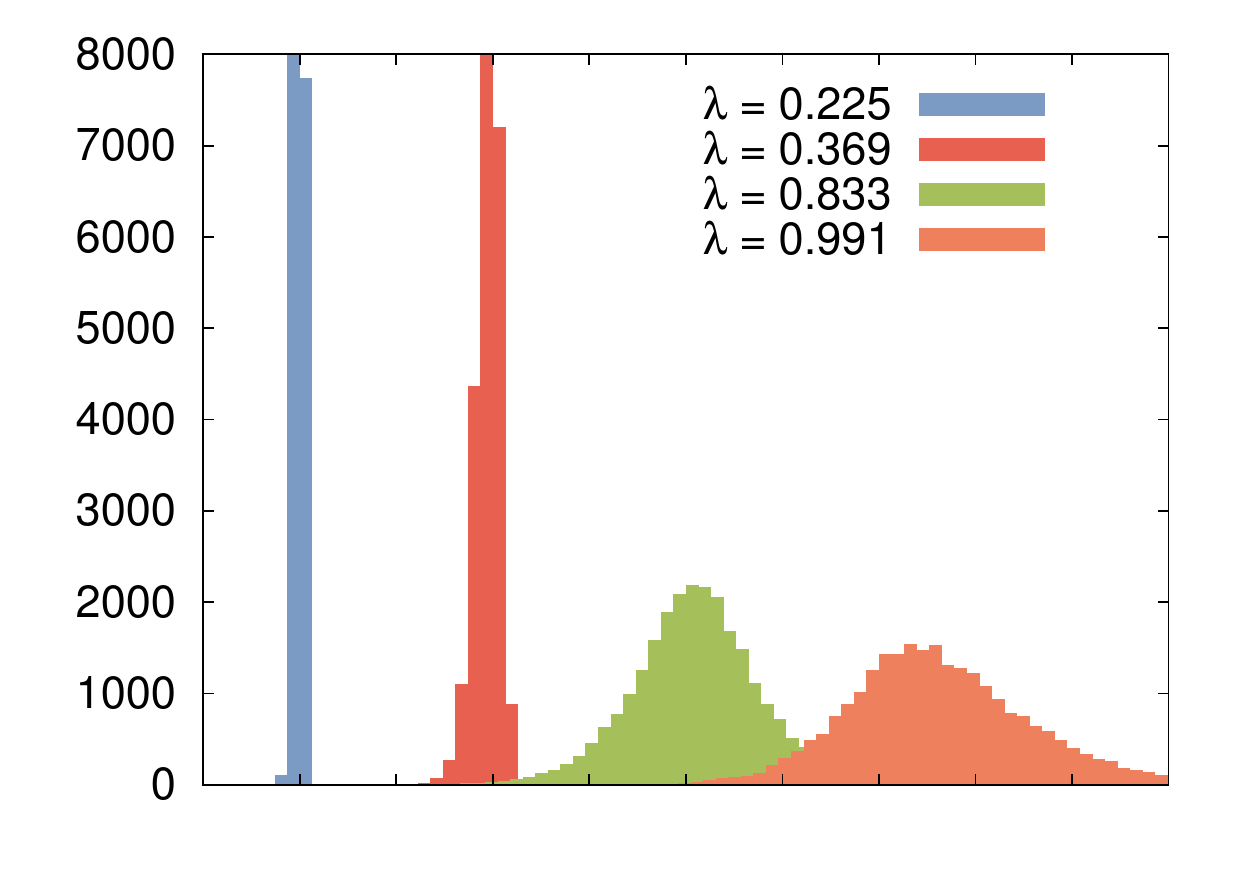}
\caption{\label{Fig:HistogramLambda}(color online) 
Histogram showing the statistical distribution of our results for $dS_2^{}/d\lambda$ for 
$L_A^{}/L=0.8$, $\lambda \simeq$ 0.225, 0.369, 0.833, 0.991, and $U/t =2.0$. 
The results for different couplings have been shifted for display purposes, but the scale 
is the same for each of them. This illustrates that low values of $\lambda$ require
less samples than larger ones in order to determine $\langle \tilde Q[\{\sigma\};\lambda] \rangle$ 
with good precision.
}
\end{figure}

In Fig.~\ref{Fig:HistogramLambda} we plot the statistical distribution of our results for the
$\lambda$ derivative at fixed region size and coupling, but varying $\lambda$. As claimed
above, the chosen parametrization requires considerably less MC samples
at low $\lambda$ than at high $\lambda$, as the width of the distributions is much smaller
for the former than for the latter.

\begin{figure}[h]
\includegraphics[width=1.0\columnwidth]{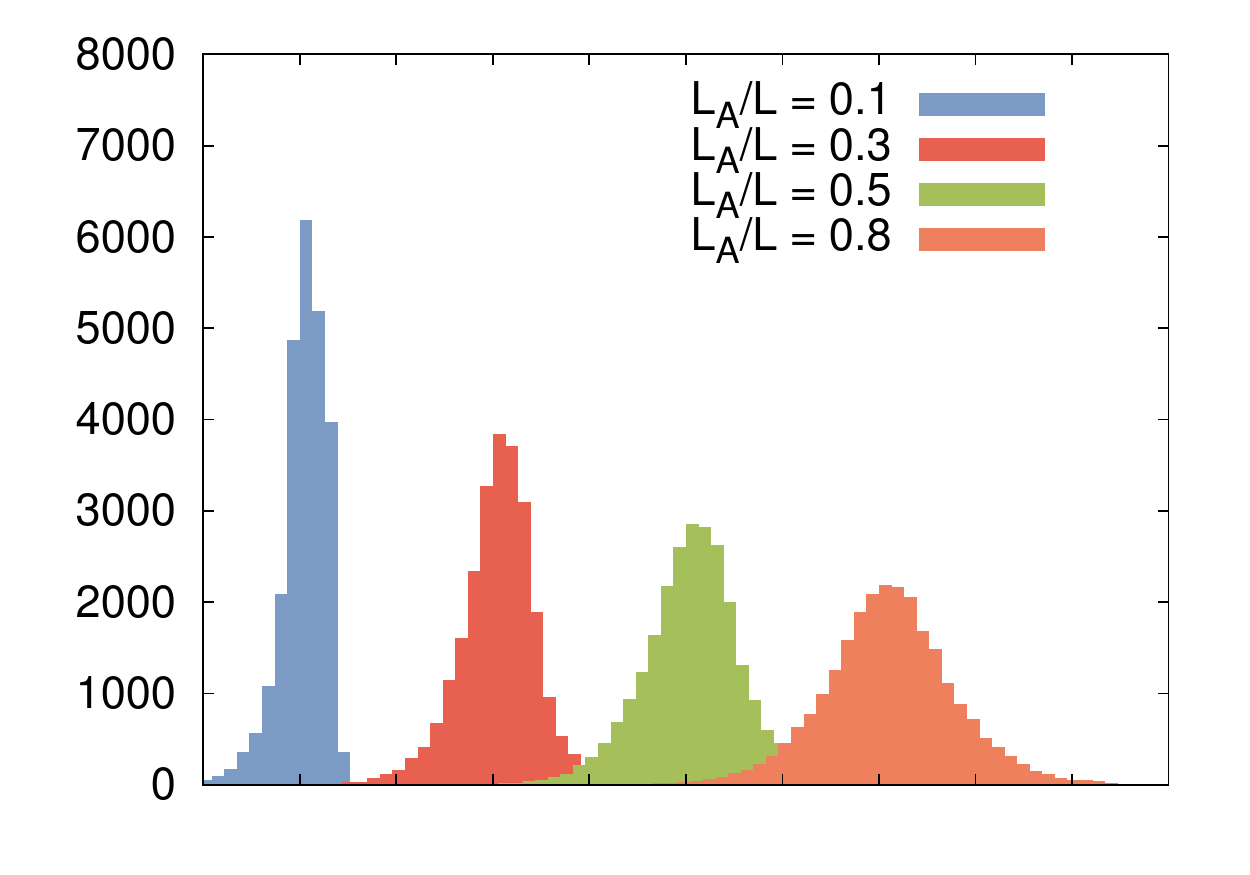}
\caption{\label{Fig:HistogramRegion}(color online) 
Histogram showing the statistical distribution of our results for $dS_2^{}/d\lambda$ for 
$L_A^{}/L=$0.1, 0.3, 0.5, 0.8, $\lambda \simeq 0.83$, and $U/t =2.0$. The results for 
different couplings have been shifted for display purposes, but the scale is the same
for each of them.
}
\end{figure}

Finally, in Fig.~\ref{Fig:HistogramRegion} we show the same distribution as above, but
as a function of subregion size. As expected, large subregions are more challenging,
but the overall shape of the distributions is very well controlled: it is close to gaussian
in that its tails decay faster than linearly in a log scale.

%%%%%%%%%%%%%%%%%%%%%%%%%%%%%%%%%%%%%%%%%%%%%%
\section{Summary and Conclusions} 

In this work, we have put forward an alternative MC approach to the calculation of the R\'enyi entanglement
entropy of many-fermion systems.
As an essential feature of our method, we compute the derivative of the entanglement entropy
with respect to an auxiliary parameter and integrate afterwards. We have shown that such a derivative can 
be computed using a MC approach without signal-to-noise issues, as the resulting expression 
yields a probability measure that does not factor across replicas and accounts for entanglement in the MC 
sampling procedure in a natural way. The subsequent numerical integration can be carried out efficiently
via Gauss-Legendre quadrature. 

As a proof of principle, we have presented results for $S^{}_2$ for the 
1D Hubbard model at half filling for different coupling strengths and compared with answers obtained by exact 
diagonalization. Our calculations show that the statistical uncertainties are well controlled, as we have shown 
in numerous plots and histograms.
Although we have not run into numerical stability issues in our tests, we anticipate that those may appear in
the form discussed in Ref.~\cite{Assaad}. Our approach is just as general as the one proposed in Ref.~\cite{Grover}. 
In particular, it can be straightforwardly generalized to finite temperature as well as to relativistic systems, in particular 
those with gauge fields such as QED and QCD, or any SU($N$) gauge theory.

%%%%%%%%%%%%%%%%%%%%%%%%%%%%%%%%%%%%%%%%%%%%%%
\acknowledgements

This work was supported in part by the U.S. National Science Foundation Nuclear Theory Program 
under Grant No. PHY1306520.

%%%%%%%%%%%%%%%%%%%%%%%%%%%%%%%%%%%%%%%%%%%%%%
% Bibliography

%%%%%%%%%%%%%%%%%%%%%%%%%%%%%%%%%%%%%%%%%%%%%%

\end{document}